\documentclass[preprint,12pt]{elsarticle} 
\usepackage{graphicx} 
\usepackage{amssymb} 
\usepackage{amsmath} 
\usepackage{amsfonts} 
\usepackage{slashed} 
\usepackage[dvipsnames]{xcolor} 

\newcommand{\be}{\begin{equation}} 
\newcommand{\ee}{\end{equation}}

\journal{Physics Letters B} 

\begin{document} 

\begin{frontmatter}

\title{Revisiting $\eta'(958)$ nuclear states} 

\author[a]{E.~Friedman} 
\author[a]{A.~Gal\corref{cor2}} 
\address[a]{Racah Institute of Physics, The Hebrew University, 9190400 
Jerusalem, Israel} 
\cortext[cor2]{corresponding author: Avraham Gal, avragal@savion.huji.ac.il}  

\begin{abstract} 

Observing $\eta'$-nuclear quasibound states requires that the $\eta'$-nuclear 
potential is both sufficiently attractive and weakly absorptive, as confirmed 
by the CBELSA/TAPS collaboration analysis of inclusive $\eta'$ production 
experiments on nuclear targets, including liquid hydrogen (LH$_2$). 
Here we present an alternative derivation of the $\eta'$-nuclear potential, 
constrained by near-threshold $pp\to pp\eta'$ and $\gamma p\to\eta' p$ 
production experiments on a free proton. The resulting $\eta'$-nuclear 
potential is weakly attractive and strongly absorptive, to the extent that 
observation of clear signals of $\eta'$-nuclear quasibound states is unlikely. 
Possible exceptions resulting from the dynamics of the nearby $I=\frac{1}{2}$ 
$J^{\pi}=(\frac{1}{2})^-$ nucleon resonance $N^{\ast}$(1895) are briefly 
discussed. 
 
\end{abstract} 


\end{frontmatter}

\section{Introduction}
\label{sec:intro}

There is considerable interest at present in meson-nuclear quasibound states, 
specifically for ${\bar K}$-$\eta$-$\eta'$ pseudoscalar mesons and 
$\omega$-$\phi$ vector mesons, all of which with masses lower than or about 
1~GeV~\cite{Metag17}. Of these meson candidates, the only meson-nuclear 
quasibound state established so far experimentally is ${\bar K}NN$, likely a 
$J^P=0^-$, $I=\frac{1}{2}$ state, bound by about 42~MeV but with a very large 
width of about 100~MeV~\cite{Yamaga20}. Experimental searches for $\eta(548)$ 
and $\eta'(958)$ quasibound nuclear states are ongoing~\cite{Bass19,Bass23} 
but so far without success. While both $\eta$ and $\eta'$ nuclear interactions 
are likely to be attractive~\cite{Metag17}, it is not clear whether or not 
they are sufficiently strong to enable quasibound states across the periodic 
table. For $\eta(548)$, considerable uncertainty arises from the model 
dependence of the poorly known near-threshold $\eta N$ scattering amplitude 
$f_{\eta N}(\sqrt{s})$ which serves as input in $\eta$-nuclear bound state 
calculations; see for example Refs.~\cite{FGM13,CFGM14}. 

For $\eta'(958)$, in contrast, most phenomenological $\eta'$-nuclear 
potentials devised for calculating quasibound states are not directly 
connected to the near-threshold $\eta' N$ scattering amplitude 
$f_{\eta' N}(\sqrt{s})$. Rather, they are constrained by fitting 
inclusive $\eta'$ photoproduction spectra taken on {\it nuclei}, 
such as $^{12}$C and $^{93}$Nb \cite{N12,N13,N16,F16,N18}. A common 
parametrization of such $\eta'$-nuclear potentials is given by~\cite{Metag17} 
\begin{equation} 
V_{\eta'}(r)=(V_0+{\rm i}W_0)\frac{\rho(r)}{\rho(0)}, 
\label{eq:Veta'} 
\end{equation} 
where $\rho(r)$ is the nuclear density. The most recent values derived for the 
strength (or depth) parameters of the real~\cite{N18} and imaginary~\cite{F16} 
parts are 
\begin{equation} 
V_0=-(44\pm 16\pm 15)~{\rm MeV},\,\,\,\,\,\,W_0=-(13\pm 3\pm 3)~{\rm MeV}. 
\label{eq:V0W0} 
\end{equation} 
Given these values, only broad and overlapping quasibound nuclear states 
are expected in $\eta'$ nuclear production spectra, as demonstrated in 
Table~\ref{tab:V0W0} for $^{12}$C and $^{40}$Ca. However, the $^{12}$C and 
$^{93}$Nb $\eta'$-photoproduction data used to extract $V_0$ and $W_0$ consist 
dominantly of $p_{\eta'}\gg k_F$ input, $k_F\approx 270$ MeV is the Fermi 
momentum at nuclear-matter density $\rho_0 \approx 0.17$~fm$^{-3}$, whereas 
construction of a near-threshold optical potential $V_{\eta'}^{\rm opt}(r)$ 
relevant to $\eta'$-nucleus bound states requires $p_{\eta'}\ll k_F$ input 
data, preferably two-body $\eta'$-nucleon data. 

\begin{table}[!htb] 
\centering 
\caption{$\eta'$ single-particle binding energies $B_{\eta'}(^{A}{\rm Z})$ 
and widths $\Gamma_{\eta'}(^{A}{\rm Z})$ in $^{12}$C and $^{40}$Ca (in MeV) 
calculated using $V_{\eta'}(r)$, Eq.~(\ref{eq:Veta'}), with $V_0=-44$~MeV, 
$W_0=-13$~MeV, and $\rho(0)\approx\rho_0=0.17$~fm$^{-3}$. Actual densities 
$\rho(r)$ are discussed in Sect.~\ref{sec:opt}.} 
\begin{tabular}{ccccc} 
\hline 
$n\ell_{\eta'}$ & $B_{\eta'}(^{12}$C) & $\Gamma_{\eta'}(^{12}$C) & 
$B_{\eta'}(^{40}$Ca) & $\Gamma_{\eta'}(^{40}$Ca) \\ 
\hline 
$1s_{\eta'}$ &  15.9  & 17.9 & 28.6 & 22.7  \\ 
$1p_{\eta'}$ & $-$1.4 & 8.0  & 16.6 & 19.7  \\ 
$1d_{\eta'}$ &   --   &  --  &  3.7 & 15.5  \\ 
\hline 
\end{tabular} 
\label{tab:V0W0} 
\end{table}  

Experiments attempting to observe $\eta'$-nuclear quasibound states have been 
limited to $^{12}$C target. A $(p,d)$ reaction using the fragment separator 
FRS at GSI was proposed by Itahashi et al.~\cite{Itahashi12}, with negative 
results published subsequently by Tanaka et al.~\cite{Tanaka16,Tanaka18}. 
Arguments in favor of trying the semi-exclusive $^{12}$C($p,dp$) reaction were 
made by Fujioka et al.~\cite{Fujioka14,Fujioka15} and very recently by Ikeno 
et al.~\cite{Ikeno24}. Another attempt, running the $^{12}$C($\gamma,p$) 
reaction in the LEPS2 beamline at SPring-8, failed to observe a clear $\eta'$ 
quasibound signal~\cite{Tomida20}. Very recently, as the present paper 
was under review, a paper by Sekiya et al.~\cite{Sekiya25} was submitted 
to PRL claiming the observation of two relatively narrow deeply bound 
$\eta'$-$^{11}$C states in the $^{12}$C$(p,d)$ reaction at GSI. Remarks on 
these findings are made here in the concluding section. 

In the present work we construct a low-energy $\eta'$-nuclear optical 
potential by using available near-threshold data of $\eta'$ production 
on a free proton. The selected data set is briefly reviewed in 
Sect.~\ref{sec:data}, and the associated $\eta'$-nuclear optical 
potential $V_{\eta'}^{\rm opt}(r)$ is discussed in Sect.~\ref{sec:opt}. 
Results of $\eta'$-nuclear quasibound state calculations are listed in 
Sect.~\ref{sec:results}. It is found that, since the imaginary part of 
$V_{\eta'}^{\rm opt}(r)$ is sizable, no clear $\eta'$ binding signals 
are expected across the periodic table. 
Concluding remarks are made in Sect.~\ref{sec:concl}.

\section{$\eta'N$ low-energy interaction parameters} 
\label{sec:data} 

Low energy $\eta'p$ scattering parameters, in particular the $\eta'p$ 
scattering length $a_{\eta'p}$, were derived from two near-threshold $\eta'$ 
production reactions on a free proton. In the first one, $a_{\eta'p}$ was 
determined from the rise of the total cross section of the near-threshold 
$\eta'$ production reaction $pp\to pp\eta'$ at COSY~\cite{COSY14}, with real 
and imaginary parts given by 
\begin{equation} 
{\rm Re}\,a_{\eta'p}=(0.00\pm 0.43)~{\rm fm} \,\,\,\,\,\, 
{\rm Im}\,a_{\eta'p}=0.37^{+0.02+0.38}_{-0.11-0.05}~{\rm fm}, 
\label{eq:cosy} 
\end{equation} 
where the systematic uncertainty of Re$\,a_{\eta'p}$ is negligible. In the 
second one, using $\gamma p\to\eta' p$ data from several near-threshold 
$\eta'$ photoproduction experiments, only the absolute value of $a_{\eta'p}$ 
was constrained~\cite{Anisovich18}: 
\begin{equation} 
|a_{\eta'p}|=(0.403\pm 0.020\pm 0.060)\,{\rm fm}, 
\label{eq:gamma1} 
\end{equation} 
with indications that the real part of $a_{\eta'p}$ is small compared 
to the imaginary part. And furthermoe, provided there is no narrow $\eta' N$ 
$s$-wave resonance near threshold, a more restrictive constraint 
arises~\cite{Anisovich18}: 
\begin{equation} 
|a_{\eta'p}|=(0.356\pm 0.012)\,{\rm fm}.  
\label{eq:gamma2} 
\end{equation} 
In fact, the nearby $I=\frac{1}{2}$ $J^{\pi}=(\frac{1}{2})^-$ $N^{\ast}$(1895) 
resonance is quite broad, $\Gamma_{1895}\approx 120$~MeV~\cite{PDG24}, thereby 
justifying this latter constraint which together with Eq.~(\ref{eq:cosy}) 
imply that Re$\,a_{\eta'p}\approx 0$ and $a_{\eta'p}$ is dominantly imaginary. 
We note that a situation where a meson-nucleon scattering length is dominantly 
imaginary is not often encountered in studies of mesic atoms and nuclei. 
For example, for $\eta$ mesons, various meson-baryon coupled-channel models 
of the broad $I=\frac{1}{2}$ $J^{\pi}=(\frac{1}{2})^-$ $N^{\ast}$(1535) 
resonance, which is peaked about 50~MeV above the $\eta N$ threshold with 
$\Gamma_{1535}\approx 150$~MeV, 
all have Re$\,a_{\eta N}\gtrsim{\rm Im}\,a_{\eta N}$~\cite{Arndt05}. 

Regarding the sign of Re$\,a_{\eta'p}$, the (10-40)\% branching ratio 
observed for $N^{\ast}(1895)\to N\eta'$ decay~\cite{PDG24} suggests that 
$N^{\ast}$(1895), projected onto the $\eta' N$ channel, is a resonance 
rather than a bound state which within our phase convention for a positive 
Im$\,a_{\eta'p}$ implies that Re$\,a_{\eta'p}$ is positive.

\section{Optical-potential methodology}
\label{sec:opt}

$\eta'$ bound states in nuclei are calculated here using $\eta'$-nuclear 
density-dependent optical potential, 
\begin{equation} 
V_{\rm opt}^{\eta'}(\rho) = 
-\frac{4\pi}{2\mu_{\eta'}}f^{(2)}_A\,b_0^A(\rho)\,\rho, 
\label{eq:V} 
\end{equation} 
with a density-dependent $\eta' N$ c.m. scattering amplitude $b_0^A(\rho)$ 
in units of fm ($\hbar=c=1$). In this expression $A$ is the mass number 
of the {\it nuclear core}, $\rho$ is a nuclear density normalized to $A$, 
$\rho_0=0.17$~fm$^{-3}$ stands for nuclear-matter density, $\mu_{\eta'}$ is 
the $\eta'$-nucleus reduced mass and $f^{(2)}_A$ is a kinematical factor 
transforming $b_0^A(\rho)$ from the $\eta' N$ center-of-mass (c.m.) system, 
to the $\eta'$-nucleus c.m. system: 
\begin{equation} 
f^{(2)}_A=1+\frac{A-1}{A}\frac{\mu_{\eta'}}{m_N}. 
\label{eq:fA} 
\end{equation} 
The density-dependent $\eta' N$ c.m. scattering amplitude $b_0^A(\rho)$ is 
given by 
\begin{equation} 
b_0^A(\rho) = \frac{{\rm Re}\,b_0}{1+(3k_F/2\pi)f^{(2)}_A{\rm Re}\,b_0} + 
{\rm i}\,{\rm Im}\,b_0,
\label{eq:b0} 
\end{equation} 
where $k_F=(3{\pi}^2\rho/2)^{1/3}$ is the Fermi momentum associated with 
local density $\rho$. The density dependence of $b_0^A(\rho)$ accounts for 
long-range Pauli correlations in $\eta' N$ in-medium multiple scatterings, 
starting at $\rho^{4/3}$~\cite{DHL71,WRW97} in a nuclear-density expansion, 
as practised in our $K^-$-atom studies~\cite{FG17}. Note that Im$\,b_0$ in 
Eq.~(\ref{eq:b0}) is not affected by Pauli correlations since all $N\eta'$ 
two-body decays proceed with c.m. momentum larger than $k_F$. The low-density 
limit of $b_0^A(\rho)$ is obtained by setting $b_0^A(\rho)\to b_0$ where 
$b_0$ is identified in the present exploratory study with the $\eta' N$ c.m. 
scattering length $a_{\eta' N}$ considered in the previous section (note that 
since $I_{\eta'}=0$, $a_{\eta' n}=a_{\eta' p}$, here denoted $a_{\eta' N}$). 
Similar forms of $V_{\rm opt}(\rho)$ were used by us recently for constructing 
the $YN$ ($Y=\Lambda,\Xi$) induced component of the $Y$-nuclear optical 
potential $V_{\rm opt}^Y(\rho)$~\cite{FG21,FG23a,FG23b,FG23c,FG25}. 

For nuclear densities $\rho(r)=\rho_p(r)+\rho_n(r)$ we used 
harmonic-oscillator type densities~\cite{Elton61} for $^{12}$C, with the same 
radial parameters for neutrons and protons, and two-parameter Fermi (2pF) 
distributions for $^{40}$Ca, normalized to $Z$ for protons and $N=A-Z$ 
for neutrons, all derived from nuclear charge distributions assembled in 
Ref.~\cite{AM13}. The corresponding r.m.s. radii follow closely values derived 
from experiment by relating proton densities $\rho_p(r)$ to charge densities 
and including the proton charge finite size and recoil effects. This approach 
is equivalent to assigning some finite range to the $\eta' N$ interaction. 
Folding reasonably chosen $\eta' N$ interaction ranges other than 
corresponding to the proton charge radius, varying the spatial form of the 
charge density, or introducing realistic differences between neutron and 
proton r.m.s. radii, made little difference.

\section{Results} 
\label{sec:results} 

Using $V_{\rm opt}^{\eta'}(\rho)$ of Eq.~(\ref{eq:V}), we report here in 
Table~\ref{tab:b0} on calculations of $\eta'$-nuclear binding energies and 
widths for $^{12}$C and $^{40}$Ca. The strength parameter $b_0$ was identified 
with the $\eta' N$ complex scattering length. Representative values of $b_0$, 
satisfying constraints deduced from $\eta'$ production experiments on 
a free proton, Eqs.~(\ref{eq:cosy})-(\ref{eq:gamma2}), were used for 
input. We first chose Im$\,a_{\eta' N}$=0.37~fm from Eq.~(\ref{eq:cosy}), 
together with either Re$\,a_{\eta' N}$=0.16~fm implied by the central value 
$|a_{\eta' p}|$=0.403~fm, Eq.~(\ref{eq:gamma1}), or Re$\,a_{\eta' N}$=0.28~fm 
implied by the upper $\pm$ value 0.466~fm there. Our second choice, 
Im$\,a_{\eta' N}$=0.25~fm which is lower by 1$\sigma$ than its central 
value of 0.37~fm, goes together with either Re$\,a_{\eta' N}$=0.25~fm which 
satisfies the central value $|a_{\eta' p}|$=0.356~fm in Eq.~(\ref{eq:gamma2}), 
or Re$\,a_{\eta' N}$=0.39~fm which satisfies $|a_{\eta' p}|$=0.466~fm, 
the upper $\pm$ value in Eq.~(\ref{eq:gamma1}). 

\begin{table}[!h] 
\centering 
\caption{$\eta'$ single-particle binding energies $B_{\eta'}(^{A}{\rm Z})$ 
and widths $\Gamma_{\eta'}(^{A}{\rm Z})$ (MeV) in $^{12}$C and $^{40}$Ca, 
calculated using $V_{\rm opt}^{\eta'}(\rho)$, Eq.~(\ref{eq:V}), for four 
values of its complex scattering length $b_0$ (fm). Potential depth values 
$D_{\eta'}=-{\rm Re}\,V_{\rm opt}^{\eta'}(\rho_0,A\to\infty)$ (MeV), with 
$\rho_0=0.17$~fm$^{-3}$, are also listed.}  
\begin{tabular}{ccccccc} 
\hline 
$b_0$ & $n\ell_{\eta'}$ & $B_{\eta'}(^{12}$C) & $\Gamma_{\eta'}(^{12}$C) & 
$B_{\eta'}(^{40}$Ca) & $\Gamma_{\eta'}(^{40}$Ca) & $D_{\eta'}$ \\ 
\hline 
0.16+i0.37 & $1s_{\eta'}$ & $-$9.73 & 34.1 & $-$0.75 & 52.9 & 11.6 \\ 
           & $1p_{\eta'}$ &   --    &  --  &    --   &  --  &      \\  
\hline 
0.28+i0.37 & $1s_{\eta'}$ & $-$4.55 & 37.7 &   5.2   & 54.3 & 18.0 \\ 
           & $1p_{\eta'}$ &   --    &  --  & $-$5.50 & 44.3 &      \\         
\hline 
0.25+i0.25 & $1s_{\eta'}$ & $-$2.75 & 21.9 &   5.1   & 35.2 & 16.5 \\ 
           & $1p_{\eta'}$ &    --   &  --  & $-$4.4  & 26.5 &      \\  
\hline 
0.39+i0.25 & $1s_{\eta'}$ &  1.76   & 25.1 &  10.7   & 36.5 & 22.6 \\ 
           & $1p_{\eta'}$ &    --   &  --  &  0.62   & 29.1 &      \\ 
\hline 
\end{tabular} 
\label{tab:b0} 
\end{table} 

The potential depth ($D_{\eta'}$) values listed in this table are smaller 
than the depth value $-V_0=44$~MeV, Eq.~(\ref{eq:V0W0}). This translates 
into smaller $\eta'$-nuclear binding energies here than those listed in 
Table~\ref{tab:V0W0}. Some of the $\eta'$ single-particle levels bound there 
are unbound here, and the presently calculated widths are considerably larger 
than width values listed there. With widths exceeding 20~MeV and reaching as 
high values as 50~MeV, there is not much to expect in searching experimentally 
for $\eta'$-nuclear quasibound states.

\section{Concluding remarks} 
\label{sec:concl} 

In this work we explored to what extent $\eta'$-nuclear quasibound states 
are sufficiently narrow to allow experimental observation. For this task 
we used a low-energy $\eta'$-nuclear density-dependent optical potential 
$V_{\rm opt}^{\eta'}(\rho)$, Eq.~(\ref{eq:V}), that has been applied by us 
quite successfully to $\Lambda$ hypernuclei~\cite{FG23b,FG23c} and $\Xi^-$ 
hypernuclei~\cite{FG21,FG23a,FG25}. The density dependence of such 
hadron-nucleus optical potential $V_{\rm opt}^h(\rho)$ accounts for 
the leading long-range Pauli correlations. For $\eta'$, the input to 
$V_{\rm opt}^{\eta'}(\rho)$ consisted of the $\eta' N$ scattering length 
$a_{\eta' N}$ which is dominated by its imaginary (absorptive) part. 
This made the $\eta'$-nuclear quasibound states calculated here, in $^{12}$C 
and in $^{40}$Ca, too broad to allow clear experimental observation. 

Future work should consider $V_{\rm opt}^{\eta'}(\rho)$ at {\it subthreshold} 
$\eta' N$ energies, as done for $\eta N$ in Refs.~\cite{FGM13,CFGM14}, where 
the subthreshold $\eta N$ scattering amplitude $a_{\eta N}(\sqrt s)$ was 
derived by following a dynamical model of $N^{\ast}$(1535) in terms of its 
main two-body decay channels, foremost $\eta N$. Given the apparent proximity 
of the $N^{\ast}$(1895) resonance to the $\eta' N$ threshold (1896-1897 MeV), 
a promising path to consider for $\eta' N$ would be a dynamical model of 
$N^{\ast}$(1895) in terms of its main decay channels $N\pi$, $N\eta$, 
$N\eta'$, $\Lambda K$, $\Sigma K$, $N\omega$, $N\rho$ and the higher-mass 
channel $N\phi$ as well. Some work in this direction was done by Bruns and 
Ciepl\'{y}~\cite{BC19}, and more recently by Sakai and Jido~\cite{SJ23}. 

Assuming that the $N^{\ast}$(1895) resonance peak is only a few MeV 
from the $\eta' N$ threshold provides a natural explanation for the 
near-vanishing of Re$\,a_{\eta' N}$. Going to subthreshold, one expects 
Re$\,a_{\eta' N}(\sqrt s)$ to rise to values of order 1 to 2 fm before 
subsiding to `normal' values of less than 1 fm, 
while Im$\,a_{\eta' N}(\sqrt s)$ quickly falls down from its maximum 
value at the resonance peak.{\footnote{This expectation derives from 
experience gained in studying the near-threshold energy dependence of 
the ${\bar K}N$ $s$-wave scattering amplitude in chiral models describing 
the $\Lambda^{\ast}$(1405) resonance, see Fig.~2 in Ref.~\cite{FGCHM17}.}}  
In a preliminary attempt to apply such a scenario to the two $\eta'$-$^{11}$C 
quasibound states with $B_{\eta'}\approx$~30~MeV and $B_{\eta'}\approx$~6~MeV 
observed very recently in the $^{12}$C($p,d$) reaction at GSI~\cite{Sekiya25}, 
we found that a subthreshold value Re$\,a_{\eta' N}(\sqrt s)\sim 3$~fm would 
come close to fit these $B_{\eta'}$ values for $1s_{\eta'}$ and $1p_{\eta'}$ 
states in $^{11}$C. Such a large Re$\,a_{\eta' N}(\sqrt s)$ would place 
severe constraints on any meson-baryon model devised for $N^{\ast}$(1895), 
posing a major challenge to any related theoretical work in the near 
future.{\footnote{Added post-publication note: discarding the $\eta'$-$^{11}$C 
deeper bound-state signal at $B_{\eta'}\approx$~30~MeV on grounds 
of poorer statistical significance than for the shallower one at 
$B_{\eta'}\approx$~6~MeV, assigning then the latter to a $1s_{\eta'}$ 
bound state, requires a smaller value of Re$\,a_{\eta' N}(\sqrt s)=(0.5
\pm 0.05)$~fm for $B_{\eta'}=(6.0\pm 1.0)$~MeV. This scenario is not 
out of reach for a suitably constructed dynamical resonance model for 
$N^{\ast}$(1895).}}

\section*{Acknowledgments}

We thank Volker Metag for drawing our attention to the recent results of 
the $^{12}$C($p,d$) experiment at GSI~\cite{Sekiya25} and its possible 
consequences. This work started as part of a project funded by the European 
Union's Horizon 2020 research \& innovation programme, grant agreement 824093.


\begin{thebibliography}{99} 

\bibitem{Metag17} V.~Metag, M.~Nanova, E.~Paryev, Prog. Part. Nucl. Phys. 97 
(2017) 199. 

\bibitem{Yamaga20} T.~Yamaga, et al. (J-PARC E15 Collab.), Phys. Rev. C 102 
(2020) 044002. 

\bibitem{Bass19} S.D.~Bass, P.~Moskal, Rev. Mod. Phys. 91 (2019) 015003. 

\bibitem{Bass23} S.D.~Bass, V.~Metag, P.~Moskal, Handbook of Nuclear Physics 
(2023) 2783 [Springer Nature, Eds: I.~Tanihata, H.~Toki, T.~Kajino]. 

\bibitem{FGM13} E.~Friedman, A.~Gal, J.~Mare\v{s}, Phys. Lett. B 725 (2013) 
334. 

\bibitem{CFGM14} A.~Ciepl\'{y}, E.~Friedman, A.~Gal, J.~Mare\v{s}, 
Nucl. Phys. A 925 (2014) 126. 

\bibitem{N12} M.~Nanova, et al. (CBELSA/TAPS Collab.), 
Phys. Lett. B 710 (2012) 600. 

\bibitem{N13} M.~Nanova, et al. (CBELSA/TAPS Collab.), 
Phys. Lett. B 727 (2013) 417. 

\bibitem{N16} M.~Nanova, et al. (CBELSA/TAPS Collab.), 
Phys. Rev. C 94 (2016) 025205.   
 
\bibitem{F16} S.~Friedrich, et al. (CBELSA/TAPS Collab.), 
Eur. Phys. J. A 52 (2016) 297.  

\bibitem{N18} M.~Nanova, et al. (CBELSA/TAPS Collab.), 
Eur. Phys. J. A 54 (2018) 182. 

\bibitem{Itahashi12} K.~Itahashi, et al., Prog. Theor. Phys. 128 (2012) 601. 

\bibitem{Tanaka16} Y.K.~Tanaka, et al., Phys. Rev. Lett. 117 (2016) 202501. 

\bibitem{Tanaka18} Y.K.~Tanaka, et al., Phys. Rev. C 97 (2018) 015202. 

\bibitem{Fujioka14} H.~Fujioka, et al., EPJ Web Conf. 66 (2014) 09006. 

\bibitem{Fujioka15} H.~Fujioka, et al., Hyperfine Interactions 234 (2015) 33. 

\bibitem{Ikeno24} N.~Ikeno, et al., arXiv:2406.06058. 

\bibitem{Tomida20} N.~Tomida, et al., Phys. Rev. Lett. 124 (2020) 202501. 

\bibitem{Sekiya25} R.~Sekiya, et al., arXiv:2509.07824v1. 

\bibitem{COSY14} E.~Czerwinski, P.~Moskal, M.~Silarski, et al., 
Phys. Rev. Lett. 113 (2014) 062004. 

\bibitem{Anisovich18} A.V.~Anisovich, et al., Phys. Lett. B 785 (2018) 626. 

\bibitem{PDG24} S.~Navas, et al. (Particle Data Group), Phys. Rev. D 110 
(2024) 030001. 

\bibitem{Arndt05} R.A.~Arndt, W.J.~Briscoe, T.W.~Morrison, I.I.~Strakovsky, 
R.L.~Workman, A.B.~Gridnev, Phys. Rev. C 72 (2005) 045202. 

\bibitem{DHL71} C.B.~Dover, J.~H\"{u}fner, R.H.~Lemmer, Ann. Phys. (NY) 66 
(1971) 248. 

\bibitem{WRW97} T.~Waas, M.~Rho, W.~Weise, Nucl. Phys. A 617 (1997) 449. 

\bibitem{FG17} E.~Friedman, A.~Gal, Nucl. Phys. A 959 (2017) 66, 
and references listed there to earlier work on $K^-$ atoms. 

\bibitem{FG21} E.~Friedman, A.~Gal, Phys. Lett. B 820 (2021) 136555. 

\bibitem{FG23a} E.~Friedman, A.~Gal, Phys. Lett. B 837 (2023) 137640. 

\bibitem{FG23b} E.~Friedman, A.~Gal, Phys. Lett. B 837 (2023) 137669. 

\bibitem{FG23c} E.~Friedman, A.~Gal, Nucl. Phys. A 1039 (2023) 122725. 

\bibitem{FG25} E.~Friedman, A.~Gal, Phys. Lett. B 868 (2025) 139728. 

\bibitem{Elton61} L.R.B~Elton, {\it Nuclear Sizes}, Oxford University Press, 
1961. 

\bibitem{AM13} I.~Angeli, K.P.~Marinova, At. Data Nucl. Data Tables 
99 (2013) 69. 

\bibitem{BC19} P.C.~Bruns, A.~Ciepl\'{y}, Nucl. Phys. A 992 (2019) 121630. 

\bibitem{SJ23} S.~Sakai, D.~Jido, Phys. Rev. C 107 (2023) 035207. 

\bibitem{FGCHM17} A.~Gal, E.~Friedman, A.~Ciepl\'{y}, J.~Hrt{\'a}nkov{\'a}, 
J.~Mare\v{s}, PoS(Hadron17) 310 (2018) 195 (DOI: doi.org/10.22323/1.310.0195). 


\end{thebibliography}
\end{document}